\documentclass[%
 reprint,
groupedaddress,
 amsmath,amssymb,
 aps,
prb,
]{revtex4-2}

\usepackage{graphicx}% Include figure files
\usepackage{braket}

\begin{document}

\title{Electronic and optical excitations of K-Sb and Na-Sb crystals}

\author{Richard Schier}
\affiliation{Carl von Ossietzky Universit\"at Oldenburg, Institute of Physics, D-26129 Oldenburg, Germany}
\author{Caterina Cocchi}
\email{caterina.cocchi@uni-oldenburg.de}
\affiliation{Carl von Ossietzky Universit\"at Oldenburg, Institute of Physics and Center for Nanoscale Dynamics (CeNaD), D-26129 Oldenburg, Germany}

\date{\today}% It is always \today, today,

\begin{abstract}
Recent advances in experimental techniques and computational methods have significantly expanded the family of alkali antimonides, a class of semiconducting materials used as photocathodes in particle accelerators, unveiling new crystal structures and stoichiometries with improved stability and quantum efficiency. This work investigates the electronic and optical properties of eight Na- and K-based alkali antimonide binary crystals with 3:1 and 1:1 alkali-to-antimony ratios, which were predicted to be stable in a recent high-throughput screening study. Employing density functional theory and many-body perturbation theory, we find that all systems exhibit direct band gaps, except for monoclinic Na$_8$Sb$_8$, which has a nearly degenerate indirect gap. Optical spectra are characterized by near-infrared absorption onsets and intense visible excitations. Our analysis highlights the significant role of electron-hole correlations, particularly in K-based compounds, leading to exciton binding energies above 100~meV and sharper absorption peaks. An in-depth analysis of the electronic contributions to the excited states provides additional insight into the role of excitonic effects. By shedding light on the fundamental properties of alkali antimonide binary crystals, our results are relevant for the design and optimization of next-generation electron sources for particle accelerators.
\end{abstract}

\maketitle

%\tableofcontents

\section{Introduction}

By complementing experimental efforts, advanced computational methods have dramatically expanded the research potential on next-generation electron sources for particle accelerators.
Since the pioneering first-principle works by Wei and Zunger in 1987~\cite{wei-zung87prb}, Tegze and Hafner in 1992~\cite{tegz-hafn92jpcm}, and Ettema and De Groot at the turn of the century~\cite{ette-degr99jpcm,ette-degr00prb,ette-degr02prb}, the study of alkali antimonides has witnessed significant advancements driven by new theoretical and numerical methods.
In the 2010s, the predictive power of density functional theory (DFT) was exploited to simulate multi-alkali antimonide crystals with various compositions and under specific structural constraints~\cite{kala+10jpcs,kala+10jpcs1,guo+14mre,murt+16bms,yala+18jmmm,khan+21ijer}.
Dedicated surface simulations have provided additional insight into the fundamental properties of these materials~\cite{schi+22prm,wang+22ssc}.  
Many-body perturbation theory (MBPT) applied on top of DFT has enabled an accurate characterization of the electronic, optical, and core-level excitations of alkali antimonides~\cite{cocc+19jpcm,amad+21jpcm,cocc-sass21micromachines,cocc20pssrrl} for insightful comparisons with experiments~\cite{cocc+19jpcm,cocc+19sr}.
Further studies on vibrational and thermal properties of these materials~\cite{zhon+21ijer,yue+22prb,liu+24prap,sant+24jpm} have contributed to disclosing their potential for thermoelectric applications~\cite{yuan+22jmcc,shar+23acsaem}. 
In parallel with calculations of known structures and compositions, high-throughput screening methods have been tailored to simulate alkali antimonides and tellurides~\cite{wang+20prm,bai+20jpcc,anto+21am,sass-cocc22jcp,schi+24ats,schi+24afm,sass-cocc24npjcm}. Data-driven schemes have not only led to an unprecedented extension of the pool of candidate materials for photocathode applications~\cite{anto+21am,schi+24afm} but also promoted the development and integration of machine learning models with atomistic simulations~\cite{anto+21am,mann24cms,sass-cocc25ats}. 

Experimental research has considerably profited from recent advances in growth and deposition techniques to obtain multi-alkali antimonide photocathodes with enhanced robustness and quantum efficiency~\cite{schu+16jap,feng+17jap,xie+17jpd,yama+18ami,schm+18prab,gald+21apl,saha+22apl,liu+22acsami}. Above all, the use of molecular beam epitaxy has led to unprecedented improvements in the synthesis of high-quality films~\cite{parz+22prl,pavl+22apl,saha+23jvstb,atam+24aplm} including new phases and stoichiometries~\cite{parz+23aplm,rozh+24prap}. The recent discovery of CsSb, which exhibits considerably higher oxidation resistance than Cs$_3$Sb~\cite{parz+23aplm}, has opened up exciting possibilities for the experimental verification of computationally predicted compounds. To fully realize this potential, reliable predictions from cutting-edge computational methods are essential.

\begin{figure}
	\centering
	\includegraphics[width=0.5\textwidth]{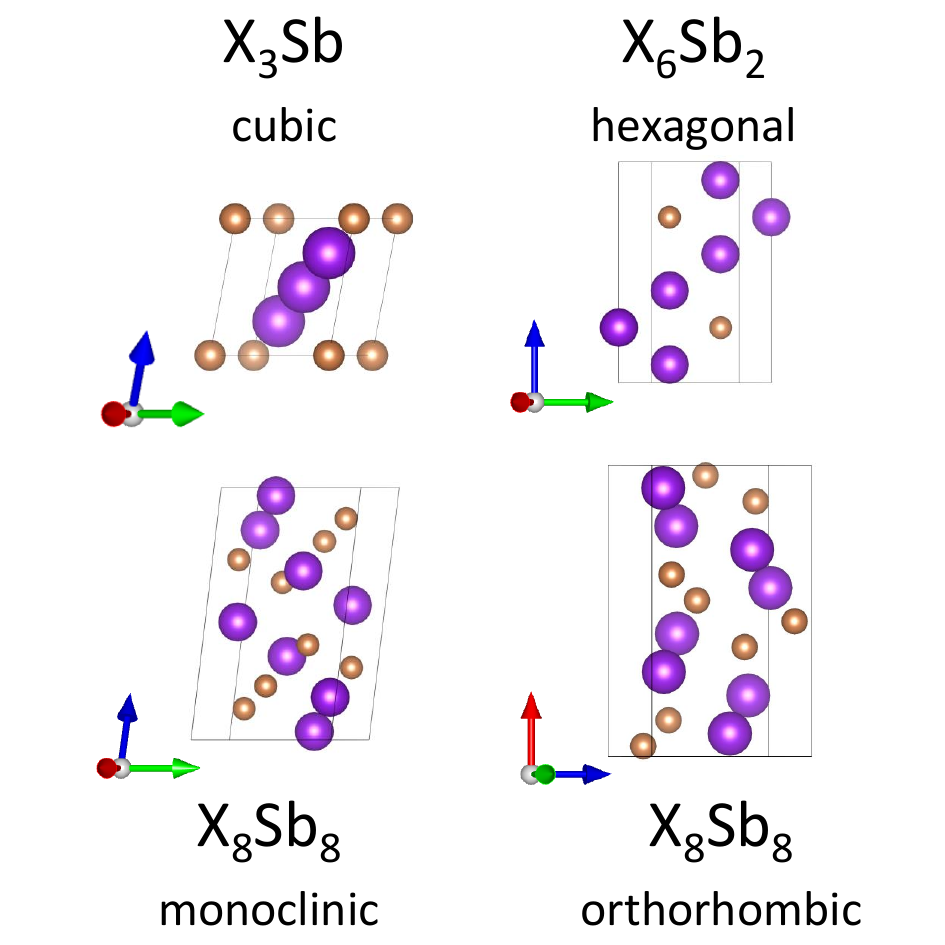}
\caption{Unit cells of the alkali antimonide crystals considered in this work. The alkali species X = Na, K are depicted in purple and Sb in bronze. Graphs produced with VESTA~\cite{momm-izun11jacr}.
}
	\label{fig:crystals}
\end{figure}

In this work based on DFT and MBPT, we investigate the electronic and optical properties of eight binary alkali antimonide semiconductors with 3:1 and 1:1 stoichiometric ratios between Na or K, and Sb (Fig.~\ref{fig:crystals}). The investigated systems are a subset of the stable structures identified in a recent high-throughput screening study~\cite{schi+24ats}. We selected these representative compounds based on their semiconducting nature, as metallic crystals were predicted with both 3:1 and 1:1 compositions~\cite{schi+24ats}. Our GW results reveal that all considered materials have a direct band gap, except for monoclinic Na$_8$Sb$_8$, which has a nearly degenerate indirect band gap. The optical spectra obtained from the solution of the Bethe-Salpeter equation exhibit onsets in the near-infrared region and intense absorption in the  visible range. Exciton binding energies are generally found below 100~meV in Na-based compounds but exceed this threshold in K-based ones, reaching 200~meV for orthorhombic K$_8$Sb$_8$. A detailed analysis of the excitations shows that electron-hole correlations lead to a redistribution of oscillator strength to lower energies and a non-trivial localization of electron-hole pairs in reciprocal space. Our findings provide valuable insights into the electronic and optical properties of binary alkali antimonide crystals, which are relevant for the development of photocathode materials for particle accelerators with optimized performance.

\section{Methodology}
\subsection{Theoretical Background}
% Theory
The results presented in this work are obtained from first principles in the framework of DFT~\cite{hohe-kohn64pr} and MBPT~\cite{onid+02rmp}. The ground-state properties are computed by solving the Kohn-Sham (KS) equation~\cite{kohn-sham65physrev},
\begin{equation}
   \hat{h} \ket{\phi_{n\text{k}}} = \epsilon_{n\text{k}} \ket{\phi_{n\text{k}}},
\end{equation}
ruled by the single-particle Hamiltonian
\begin{equation}
    \hat{h} = \hat{T} + \hat{V}^{\text{eff}} = \hat{T} + \hat{V}^{\text{ext}} + \hat{V}^{\text{H}} + \hat{V}^{\text{xc}}.
    \label{eq:KS-Ham}
\end{equation}
In addition to the kinetic energy operator $\hat{T}$, the KS Hamiltonian in Eq.~\eqref{eq:KS-Ham} contains the effective potential $\hat{V}^{\text{eff}}$, which consists of the external potential $\hat{V}^{\text{ext}}$, accounting for the electron-nuclear attraction, the Hartree potential $\hat{V}^{\text{H}}$, and the exchange-correlation potential $\hat{V}^{\text{xc}}$, embedding all the electronic interaction beyond the classical picture. Since the exact form of $\hat{V}^{\text{xc}}$ is not known, approximations must be taken for this term, determining the accuracy of the DFT calculations.

To access excited-state properties, DFT results are used as a starting point for the GW approximation~\cite{hedi65:physrev} in the single-shot G$_0$W$_0$ approach to calculate the electronic self-energy
\begin{equation}
\Sigma (r, r', \omega) = \frac{i}{2\pi} \int G_0 (r,r',\omega + \omega') W_0(r, r', \omega') e^{i \omega' \eta} d\omega',
\label{eq:Sigma}
\end{equation}
where $G_0$ is the single-particle Green's function and $W_0 = \epsilon^{-1} v_c$ is the screened Coulomb potential with $\epsilon^{-1}$ being the frequency-dependent dielectric function of the material and $v_c$ the bare Coulomb potential. The self-energy computed from Eq.~\eqref{eq:Sigma} enters the quasi-particle (QP) equation
\begin{equation}
\epsilon_{n\text{k}}^{\text{QP}} = \epsilon_{n\text{k}} + Z_{n\text{k}} [\Re\Sigma_{n\text{k}} (\epsilon_{n\text{k}}) - V_{n\text{k}}^{\text{xc}}],
\label{eq:QP}
\end{equation}
where the renormalization factor $Z_{n\text{k}}$ accounts for the energy-dependence of the self-energy.

To access optical properties, including bound electron-hole pairs, we solve the Bethe-Salpeter equation (BSE)~\cite{salp-beth51physrev} as an effective eigenvalue problem,
\begin{equation}
\hat{H}^{\text{BSE}} \ket{A^{\lambda}} = E^{\lambda}\ket{A^{\lambda}},
\label{eq:BSE}
\end{equation}
delivering eigenvalues $E^{\lambda}$ and eigenvectors $\ket{A^{\lambda}}$ upon diagonalization.
The BSE Hamiltonian in Eq.~\eqref{eq:BSE} is composed of three terms: 
\begin{equation}
\hat{H}^{\text{BSE}} = \hat{H}^{\text{diag}} + 2 \hat{H}^{\text{x}} + \hat{H}^{\text{c}}.
\label{eq:H_BSE}
\end{equation}
The diagonal term $\hat{H}^{\text{diag}}$ describes vertical electronic transitions, the exchange term $\hat{H}^{\text{x}}$, multiplied by 2 in spin-degenerate systems, accounts for the repulsive exchange interaction between the fermionic electron-hole pairs, and the direct term $\hat{H}^{\text{c}}$ includes the attractive electron-hole Coulomb interaction.
The solutions of Eq.~\eqref{eq:BSE} can be used to compute optical spectra in the optical limit expressed by the imaginary part of the macroscopic dielectric function
\begin{equation}
 \text{Im } \epsilon_M = \frac{8 \pi^2}{\Omega} \sum_{\lambda} |\mathbf{t}^{\lambda}|^2 \delta(\omega - E^{\lambda}),
\end{equation}
where, in addition to the BSE eigenvalues $E^{\lambda}$, we identify the unit cell  volume $\Omega$ and the transition coefficients $\mathbf{t}^{\lambda}$, including the momentum matrix elements between valence ($v$) and conduction states ($c$) weighted by the BSE eigenvectors:
\begin{equation}\label{eqn:transitioncoeffs}
\mathbf{t}^\lambda=\sum_{v c \text{k}} A_{v c \text{k}}^\lambda \frac{\langle v \text{k}|\hat{\mathbf{p}}| c \text{k}\rangle}{\epsilon_{c \text{k}}^{\rm QP}-\epsilon_{v \text{k}}^{\rm QP}}.
\end{equation}
The coefficients $A_{v c \text{k}}^\lambda$ contain information about the single-particle composition of each excitation which can be visualized on top of the QP band structure using the so-called \textit{exciton weights} for holes
\begin{equation}
w^{\lambda}_{v\mathbf{k}} = \sum_c |A^{\lambda}_{vc\mathbf{k}}|^2,
\label{eq:w_h}
\end{equation}
and electrons
\begin{equation}
 w^{\lambda}_{c\mathbf{k}} = \sum_v |A^{\lambda}_{vc\mathbf{k}}|^2.
\label{eq:w_e}
\end{equation}
% 

% Computational Parameters
\subsection{Computational Details}

All calculations are performed with the code \texttt{exciting}~\cite{gula+14iop,vorw+19es} implementing the all-electron full-potential formalism with a mix of linearised augmented plane waves and local orbitals. Muffin-tin radii of 2.0~bohr are used for both alkali species, K and Na, while a value of 2.2~bohr is chosen for Sb. The cutoff for the plane-wave part of the basis is set to 8.5 and the adopted k-meshes are reported in Table~\ref{tab:k-mesh}. The PBE-sol functional~\cite{perd+08physrevl} approximates $\hat{V}^{\text{xc}}$. The crystal structures are taken from previous work~\cite{schi+24ats} without further optimization. Lattice parameters and angles are provided in the Supplemental Material (SM), Table~S1.

\begin{table}[h]
\centering
\caption{Summary of the k-point meshes adopted for DFT, G$_0$W$_0$, and BSE calculations on the alkali antimonides investigated in this work. In the stoichiometry, X = Na, K. The labels ``o" and ``m" stand for orthorhombic and monoclinic symmetry, respectively.}
\label{tab:k-mesh}
\begin{tabular}{c|c|c|c|c}
\hline
Stoichiometry & Symmetry & DFT & G$_0$W$_0$ & BSE \\
\hline
X$_3$Sb & cubic & 10 $\times$ 10 $\times$ 10 & 6 $\times$ 6 $\times$ 6 & 8 $\times$ 8 $\times$ 8 \\
X$_6$Sb$_2$ & hexagonal & 8 $\times$ 8 $\times$ 4 & 4 $\times$ 4 $\times$ 2 & 8 $\times$ 8 $\times$ 4 \\
X$_8$Sb$_8$ & o/m & 8 $\times$ 8 $\times$ 4 & 4 $\times$ 4 $\times$ 2 & 6 $\times$ 6 $\times$ 3 \\
\hline
\end{tabular}
\end{table}

In the G$_0$W$_0$ runs~\cite{nabo+16prb}, the dielectric screening is calculated in the random-phase approximation (RPA) including 200 empty states. 
The BSE calculations are performed in the Tamm-Dancoff approximation with 100 empty states to compute static screening in the RPA. A local-field cutoff of 1.5~Ha is chosen after careful convergence tests (see SM, Fig.~S1a). To solve the BSE, 4 (14) occupied (unoccupied) states are included for X$_3$Sb, 6 (12) occupied (unoccupied) states for X$_6$Sb$_2$, and 10 (16) occupied (unoccupied) states for X$_8$Sb$_8$, where X = K, Na. The k-grids adopted at this stage are summarized in Table~\ref{tab:k-mesh} and ensure appropriate convergence of the lowest-energy region of the spectra, see SM, Fig.~S1b.

\section{Results and Discussion}

The systems investigated in this work are eight binary alkali antimonide crystals with Na and K cations, see Fig.~\ref{fig:crystals}. 

The cubic phases K$_3$Sb and Na$_3$Sb are the established experimental compounds~\cite{ebin-taka73prb}. The hexagonal structures K$_6$Sb$_2$ and Na$_6$Sb$_2$ have been identified as stable in a recent high-throughput screening study based on DFT~\cite{schi+24afm}. Although their experimental verification is still missing, they are relevant in light of the hexagonal lattice characterizing the experimentally stable phase of the ternary compound NaK$_2$Sb~\cite{mcca60jpcs}. Among the isostoichiometric crystals, we examine K$_8$Sb$_8$ and Na$_8$Sb$_8$ in both their monoclinic and orthorhombic phases. An earlier \textit{ab initio} study indicated monoclinic KSb as the most stable polymorph with this stoichiometry~\cite{seif-hafn99prb}. However, it is possible that both monoclinic and orthorhombic phases can coexist as recently demonstrated for CsSb~\cite{parz+23aplm}.
We focus on two particularly relevant compositions, namely 3:1 and 1:1 alkali-to-antimony ratios which have been identified as stable compounds in a recent high-throughput screening study based on DFT~\cite{schi+24ats}. The 3:1 stoichiometry corresponds to the global minimum of the formation energy of Na-Sb crystals, lying approximately 200~meV below the most stable isostoichiometric compound~\cite{schi+24ats}. In the K-Sb  configurational space, the global minimum for formation energy was found for monoclinic K$_5$Sb$_4$. However, the most favorable 1:1 and 3:1 crystals exhibit formation energies only a few tens of meV higher than K$_5$Sb$_4$~\cite{schi+24ats}.

\subsection{Electronic Properties}
\label{ssec:electronic}
\begin{figure*}
	\centering
	\includegraphics[width=1.0\textwidth]{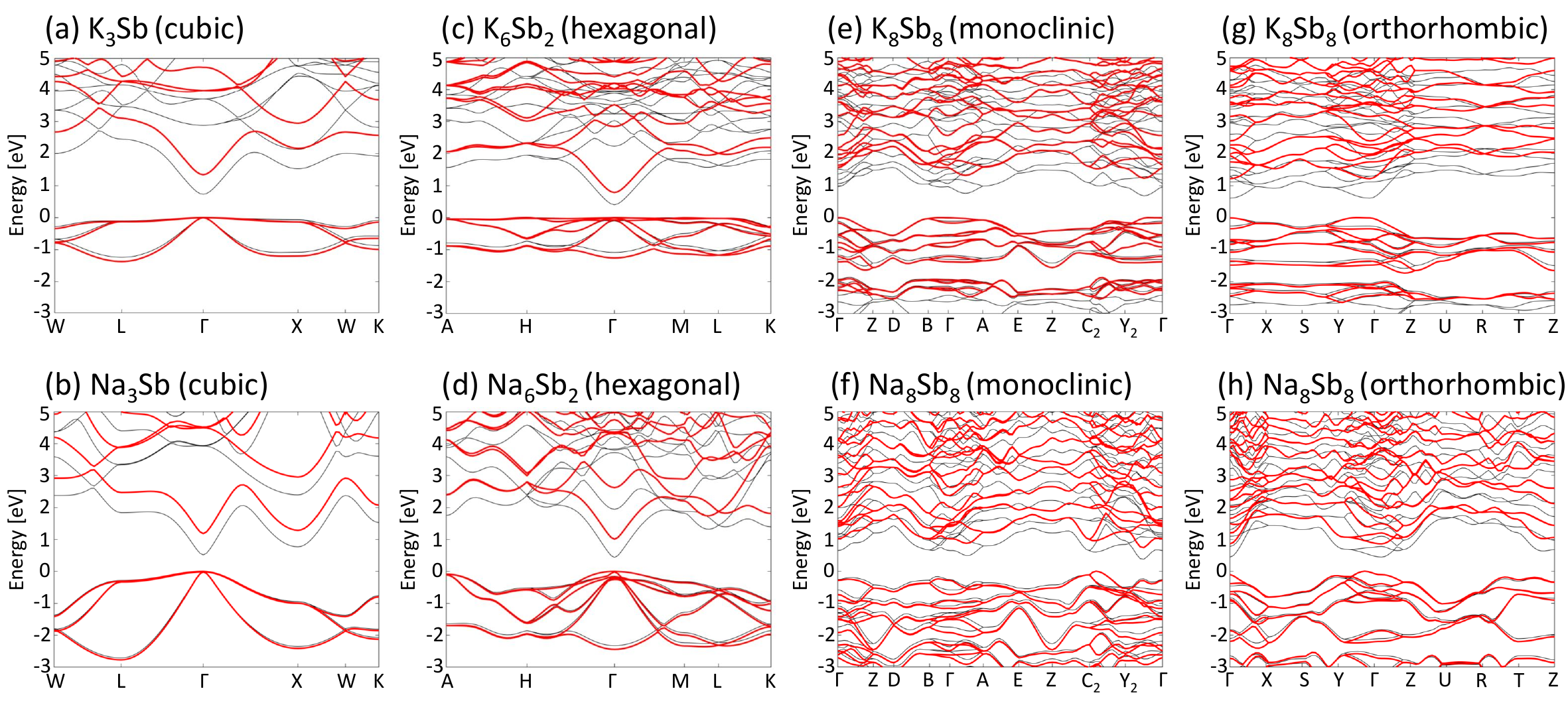}
\caption{Band structures of the alkali antimonide crystals considered in this work calculated from DFT (gray) and G$_0$W$_0$ (red) and aligned at the valence band maximum set to 0~eV. Each panel reports the chemical formula of the compound and its crystal symmetry in parenthesis.
}
	\label{fig:bandstructures}
\end{figure*}

We start our analysis by discussing the electronic band structure of the considered alkali antimonide crystals obtained from DFT and G$_0$W$_0$, see Fig.~\ref{fig:bandstructures} and Table~\ref{table:gaps}.
The projected densities of states are provided in the SM, Figs.~S2 and S3.
Cubic K$_3$Sb and Na$_3$Sb exhibit a direct QP gap at $\Gamma$ of magnitude 1.34~eV and 1.18~eV, respectively. In both cases, the self-energy contribution enhances the DFT gap by about 0.6~eV without altering the band dispersion. However, a detailed inspection of Fig.~\ref{fig:bandstructures}a-b reveals subtle differences in the QP correction at different high-symmetry points. While  the conduction-band minimum (CBm) remains at the zone center for both materials, the energy difference between $\Gamma$ and X obtained from G$_0$W$_0$ is reduced in Na$_3$Sb (see Fig.~\ref{fig:bandstructures}b). The QP correction also influences the manifold of occupied states visualized in Fig.~\ref{fig:bandstructures}a-b. In K$_3$Sb, the self-energy contribution broadens the valence band by a few tens of meV, particularly affecting the deepest state (see Fig.~\ref{fig:bandstructures}a). In contrast, these effects are less pronounced in Na$_3$Sb, where the valence bands are intrinsically more dispersive, see Fig.~\ref{fig:bandstructures}b, although in both materials the uppermost occupied states are dominated by Sb $p$-contributions (see Fig.~S2). The decreasing width of the highest valence band manifold as the size of the alkali species increases is consistent with the results obtained for cubic Cs$_3$Sb at the same level of theory~\cite{cocc+19sr}. This characteristic can be related to the variations in the band gaps of the binary alkali antimonide crystals, see Table~\ref{table:gaps}.

The hexagonal crystals with 3:1 stoichiometry, K$_6$Sb$_2$ and Na$_6$Sb$_2$, exhibit smaller energy gaps than their cubic counterparts: 0.79~eV for K$_6$Sb$_2$ and 1.01~eV for Na$_6$Sb$_2$, see Table~\ref{table:gaps}. Remarkably, the QP correction in these systems is again around 0.6~eV. The overall band dispersion is the same in the DFT and G$_0$W$_0$ results and both materials retain their direct band gap at $\Gamma$, see Fig.~\ref{fig:bandstructures}c-d. The valence-band manifold of K$_6$Sb$_2$ remains relatively flat, spanning approximately 1~eV, similar to K$_3$Sb, even with the self-energy contribution. Conversely, the uppermost occupied bands of Na$_6$Sb$_2$ exhibit larger dispersion, similar to Na$_3$Sb. In both K$_6$Sb$_2$ and Na$_6$Sb$_2$, the QP correction to the valence region is of the order of a few tens of meV.

\begin{table}[h]
\centering
\caption{DFT and QP gaps ($E^{DFT}_{gap}$ and $E^{QP}_{gap}$, respectively) as well as the energy of the first excitation $E_1$ computed from the solution of the BSE and the corresponding exciton binding energy $E_b = E^{QP}_{gap} - E_1$ for the alkali antimonide crystals considered in this work. Energies associated with optically dark lowest-energy excitations are reported in parentheses. All energy values are in eV.}
\label{table:gaps}
\begin{tabular}{c|c||c|c||c|c}
\hline
Stoichiometry & Symmetry & $E^{DFT}_{gap}$ & $E^{QP}_{gap}$ & $E_1$ & $E_b$ \\
\hline
K$_3$Sb & cubic & 0.73 & 1.34 & 1.23 & 0.12 \\
K$_6$Sb$_2$ & hexagonal & 0.40 & 0.79 & 0.68 & 0.11 \\
K$_8$Sb$_8$ & monoclinic & 0.67 & 1.29 & (1.17) & (0.12) \\
K$_8$Sb$_8$ & orthorhombic & 0.61 & 1.23 & 1.03 & 0.20 \\ \hline
Na$_3$Sb  & cubic & 0.52 & 1.18 & 1.11 & 0.07 \\
Na$_6$Sb$_2$ & hexagonal & 0.44 & 1.01 & 0.92 & 0.09 \\
Na$_8$Sb$_8$ & monoclinic & 0.38 & 1.05 & (0.98) & (0.07) \\
Na$_8$Sb$_8$ & orthorhombic & 0.45 & 0.90 & (0.81) & (0.09) \\
\hline
\end{tabular}
\end{table}

The band structures of the four semiconducting crystals with 1:1 stoichiometry are characterized by a large number of states in the region displayed in Fig.~\ref{fig:bandstructures}e-h. This behavior is easily understood by considering that the unit cells of these materials host a total of 16 atoms in contrast with the 4 atoms of the cubic phases and the 8 atoms of the hexagonal ones, see Fig.~\ref{fig:crystals}. The monoclinic crystal K$_8$Sb$_8$ has a direct QP gap of 1.29~eV at the valence-band maximum (VBM), which is found between Y$_2$ and $\Gamma$, see Fig.~\ref{fig:bandstructures}e. This value is approximately twice the one delivered by DFT (Table~\ref{table:gaps}). On the other hand, monoclinic Na$_8$Sb$_8$ is an indirect band-gap semiconductor with the VBM close to C$_2$ and the CBm between Y$_2$ and $\Gamma$. The QP gap is 1.05~eV, resulting from a self-energy contribution of 0.67~eV added to the DFT result, see Fig.~\ref{fig:bandstructures}f and Table~\ref{table:gaps}. The dispersion of the lowest unoccupied band characterizing the band structures of the cubic and hexagonal crystals is less pronounced here. This characteristic is explained by the larger contributions from the Sb atoms in the conduction region of these isostoichiometric phases~\cite{schi+24ats}. In the valence, the relatively narrow uppermost manifold seen in the band structure of the K-containing cubic and hexagonal crystals can be recognized also in monoclinic K$_8$Sb$_8$. However, the different stoichiometry is responsible for the presence of another set of bands with a predominant Sb character approximately 0.5~eV below the highest one, see Fig.~\ref{fig:bandstructures}e. Similarly, a continuum of bands appears in the valence region of Na$_8$Sb$_8$ (Fig.~\ref{fig:bandstructures}f), in contrast to the gap visible in the bottom part of Fig.~\ref{fig:bandstructures}b and Fig.~\ref{fig:bandstructures}d.

The orthorhombic phases on K$_8$Sb$_8$ and Na$_8$Sb$_8$ are characterized by direct QP gaps at $\Gamma$ of magnitude 1.23~eV and 0.90~eV, respectively, see Fig.~\ref{fig:bandstructures}g-h and Table~\ref{table:gaps}. Notably, the self-energy contribution in Na$_8$Sb$_8$ is only 0.45~eV, in contrast to 0.62~eV obtained for K$_8$Sb$_8$. Again, the QP correction does not qualitatively alter the band structure reproduced by DFT, with the large number of atoms in the unit cell and the equal proportion of alkali and antimony species therein being responsible for the large density of states with moderate dispersion, see Fig.~\ref{fig:bandstructures}g-h. 

The analysis of the electronic structure of these eight alkali antimonide crystals provides some important indications: (i) all materials exhibit QP gap energies below the range of visible radiation; (ii) the K-containing crystals are characterized by systematically larger band gaps than their Na-based counterparts except for the hexagonal crystals with 3:1 alkali-antimony ratio; (iii) all alkali antimonides considered in this study have direct band gaps apart from monoclinic Na$_8$Sb$_8$, where, however, the direct gap is only 220~meV higher than the fundamental (indirect) one. With this information, we can move on to inspecting the optical properties.

%\begin{table}[h]
%\begin{tabular}{|l|r|r|r|r|}
%\hline
%                & DFT gap  & QP gap  & $E$ (E1)& \hspace{0.3cm} $E_b$ \\ \hline
%K$_3$Sb         & 0.73     & 1.34    & 1.23    & 0.12   \\ \hline
%K$_6$Sb$_2$     & 0.40     & 0.79    & 0.68    & 0.11   \\ \hline
%K$_8$Sb$_8$(m)  & 0.67     & 1.29    & 1.17    & 0.12   \\ \hline
%K$_8$Sb$_8$(o)  & 0.61     & 1.23    & 1.03    & 0.20   \\ \hline
%Na$_3$Sb        & 0.52     & 1.18    & 1.11    & 0.07   \\ \hline
%Na$_6$Sb$_2$    & 0.44     & 1.01    & 0.92    & 0.09   \\ \hline
%Na$_8$Sb$_8$(m) & 0.38     & 1.05    & 0.98    & 0.07   \\ \hline
%Na$_8$Sb$_8$(o) & 0.45     & 0.90    & 0.81    & 0.09   \\ \hline
%\end{tabular}
%\caption{Gaps from DFT and G$_0$W$_0$ as well as the energy of the first excitation from BSE $E$ (E1) together with its excitonic binding energy $E_b$. All values are in eV.}
%\label{table:gaps}
%\end{table}

\subsection{Optical Properties}
\begin{figure*}
	\centering
	\includegraphics[width=1.0\textwidth]{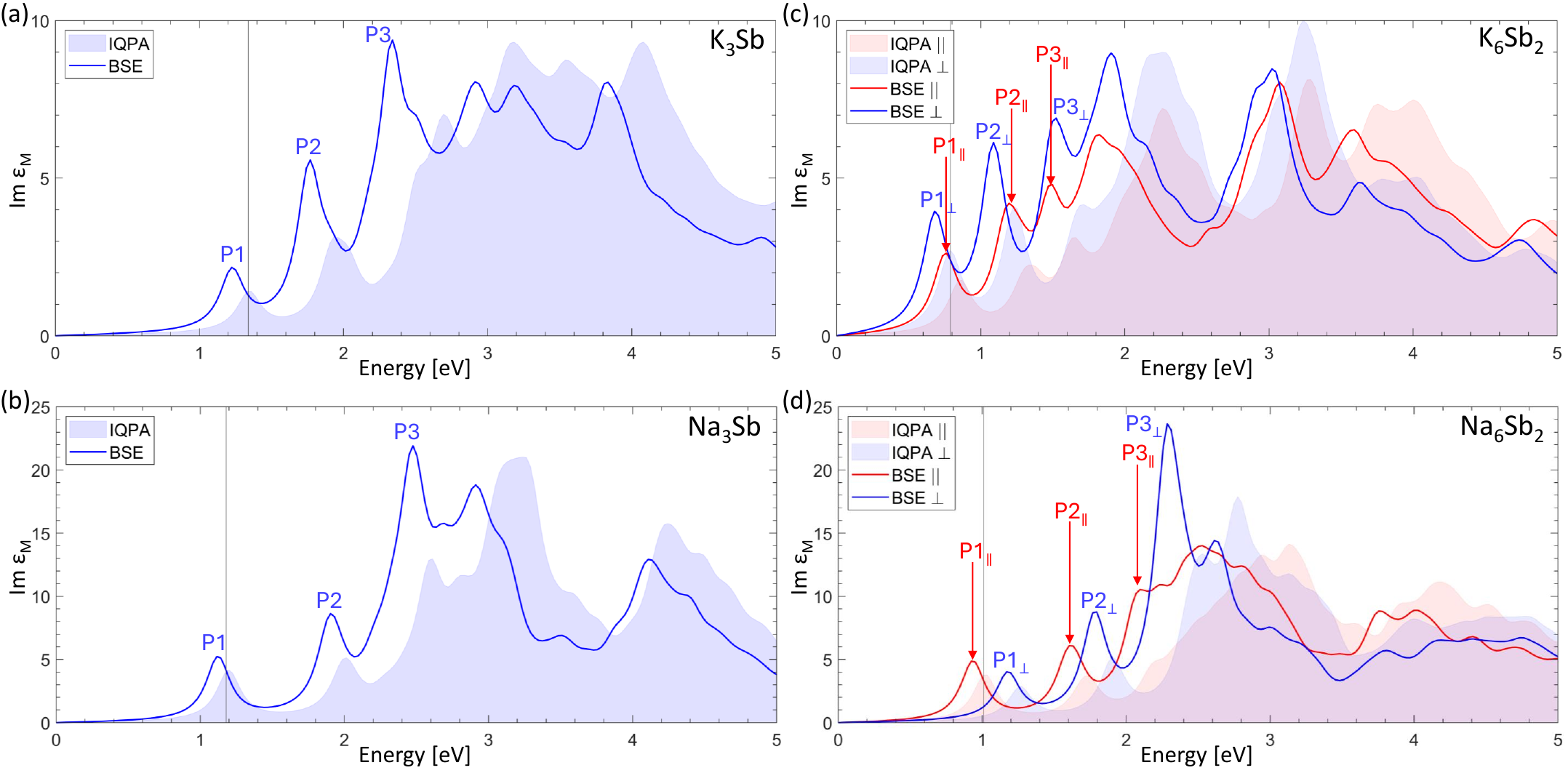}
\caption{Optical spectra of the cubic crystals (a) K$_3$Sb and (b) Na$_3$Sb as well as of the hexagonal crystals (c) K$_6$Sb$_2$ and (d) Na$_6$Sb$_2$. The solid curves and the shaded areas represent the diagonal components of the imaginary part of the macroscopic dielectric tensor calculated from the solution of the BSE and in the IQPA, respectively. For the hexagonal phases, both in-plane ($\parallel$) and out-of-plane ($\perp$) components are plotted. A Lorentzian broadening of 100~meV is applied to visualize the spectra. In all plots, the QP gap is marked by a vertical bar.}
	\label{fig:os_31}
\end{figure*}

The analysis of the electronic properties of the alkali antimonide crystals presented in Sec.~\ref{ssec:electronic} has shown intrinsic differences between materials with 3:1 and 1:1 stoichiometric ratios. For this reason, we split the investigation of the optical properties focusing first on the cubic and hexagonal compounds with chemical formulas X$_3$Sb and X$_6$Sb$_2$, respectively (X = K, Na). Subsequently, we will examine the results obtained for the monoclinic and orthorhombic isostoichiometric crystals.

As expected from the trends reported for the QP gaps in Table~\ref{table:gaps}, the cubic crystals K$_3$Sb and Na$_3$Sb are characterized by an absorption onset in the near-infrared region, see Fig.~\ref{fig:os_31}a-b. In both materials, the lowest-energy excitation is bright and gives rise to peak P1, centered at 1.23~eV and 1.11~eV in the spectrum of K$_3$Sb and Na$_3$Sb, respectively. The second maximum, P2, appears at visible frequencies, about 0.5~eV higher in energy. In K$_3$Sb, the oscillator strength of P2 is almost three times larger than the one of P1, while for Na$_3$Sb, this difference is reduced by a factor of 2. The most intense maximum visualized in Fig.~\ref{fig:os_31}a-b is associated with P3, appearing at about 2.3~eV in both spectra. At higher energies, the spectrum of K$_3$Sb exhibits an intense continuum of states entering the ultraviolet (UV) region, see Fig.~\ref{fig:os_31}a. On the other hand, in Na$_3$Sb, we find a significant absorption drop starting at the edge between the visible and the UV band ($\sim$3.2~eV) up to about 4~eV, where another maximum appears in Fig.~\ref{fig:os_31}b.

To validate our computational results, we compare the calculated spectra of K$_3$Sb and Na$_3$Sb with available experimental data. 
According to the measurements by Ebina and Takahashi~\cite{ebin-taka72josa,ebin-taka73prb} and Spicer~\cite{spic58pr}, the imaginary part of the dielectric function of K$_3$Sb has an onset around 1.2~eV and a prominent peak near 2.4~eV. These findings are consistent with our results for P1 and P3 reported in Fig.~\ref{fig:os_31}a, while P2 can be associated with a shoulder observed between the onset and the primary peak in the experimental data. The measurements reveal a second intense peak in Im$\varepsilon_M$ near 3.5~eV~\cite{ebin-taka73prb}. While a corresponding distinct feature cannot be identified in Fig.~\ref{fig:os_31}a, our computational results predict a region of intense absorption in the same window. 
A similar good agreement with experiments is obtained also for our calculations on Na$_3$Sb. The measurements indicate an onset at approximately 1~eV and an intense absorption maximum between 2.5 and 3~eV~\cite{spic58pr,ebin-taka73prb}. These features match well with P1 and P3 in Fig.~\ref{fig:os_31}b, respectively, while P2, appearing as a well-defined maximum in our result, is assimilated in the steep absorption curve recorded in the experiments. 

To assess the influence of electron-hole correlations, we overlaid the BSE spectrum (solid lines in Fig.~\ref{fig:os_31}a-b) with the result obtained in the independent QP approximation (IQPA, shaded areas). In the IQPA, the BSE Hamiltonian (Eq.~\ref{eq:BSE}) is truncated to its diagonal term, $H^{BSE} = H^{diag}$, neglecting both the direct and exchange Coulomb interactions. As expected, the inclusion of electron-hole correlations in the BSE calculations leads to a redshift of the absorption peaks, due to the prevalence of the electron-hole Coulomb attraction over the repulsive exchange term. However, the magnitude of this redshift varies across different excitations, indicating a heterogeneous influence of the excitonic effects in different spectral ranges. By evaluating the exciton binding energy as the difference between excitation energy obtained in the IQPA and from the solution of the BSE~\cite{cocc+19jpcm,amad+21jpcm,cocc-sass21micromachines}, 
\begin{equation}
    E^{\lambda}_b = E^{\lambda}_{IQPA} - E^{\lambda}_{BSE},
    \label{eq:Eb}
\end{equation}
we obtain a value for P1 of 0.12~eV in K$_3$Sb and 0.07~eV in Na$_3$Sb, see Table~\ref{table:gaps}. P2 exhibits a larger binding energy than P1 in both cubic crystals: in K$_3$Sb, $E_b^{P2}=0.19$~eV while in Na$_3$Sb, $E_b^{P2}=0.11$~eV. The same estimate can be applied to P3 although in K$_3$Sb, the spectral shape changes when excitonic effects are included (BSE) or excluded (IQPA), see Fig.~\ref{fig:os_31}a. Estimating the binding energy of P3 in this material from the nearest maximum in the IQPA spectrum, corresponding to a shoulder, we obtain a value of 0.19~eV. In Na$_3$Sb, $E_b^{P3}=0.13$~eV. We do not continue with this analysis for the other maxima since they lie above the energy region of interest for photocathode applications. 
Notably, the binding energies obtained for cubic K$_3$Sb and Na$_3$Sb are of the same order as those characterizing not only cubic Cs$_3$Sb~\cite{cocc-sass21micromachines} but also the ternary crystals CsK$_2$Sb~\cite{cocc+19jpcm}, NaK$_2$Sb, and Na$_2$KSb~\cite{amad+21jpcm}.

\begin{figure}
	\centering
	\includegraphics[width=0.5\textwidth]{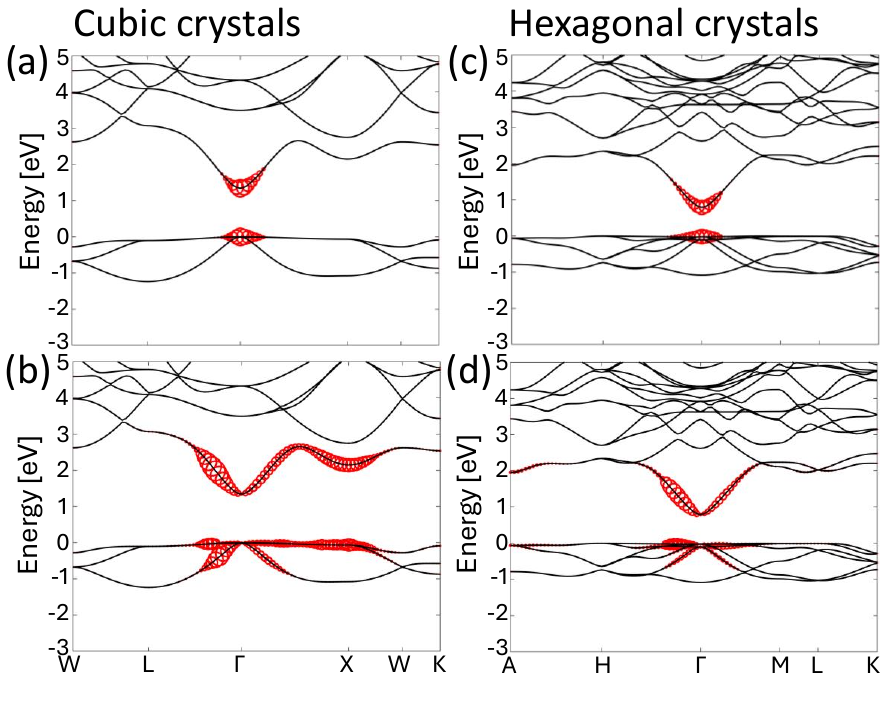}
\caption{Band-structure distribution of selected exciton weights, indicated by red dots of varying size, of (a) P1 and (b) P3 of K$_3$Sb as well as for (c) P1$_{\perp}$ and (d) P3$_{\parallel}$ of K$_6$Sb$_2$. }
	\label{fig:exc_31}
\end{figure}

The excitonic character of the lowest-energy peaks in the spectra of K$_3$Sb and Na$_3$Sb can be further analyzed from the electronic band contributions and their distribution in k-space. As expected from the electronic-structure analysis reported in Fig.~\ref{fig:bandstructures}a-b and Ref.~\cite{schi+24ats}, the lowest-energy excitation, P1, originates from a vertical transition at the $\Gamma$ point between the VBM, dominated by Sb $p$-contributions~\cite{ette-degr99jpcm,kala+10jpcs}, and the CBm with mostly K $s$-nature~\cite{ette-degr99jpcm,kala+10jpcs} (see Fig.~S2a in the SM), leading to a delocalized Wannier exciton in real space typical of in inorganic semiconductors~\cite{fox10book}. This character is mirrored by the strong localization in k-space illustrated in Fig.~\ref{fig:exc_31}a for P1 in K$_3$Sb. 
The result obtained for P1 in Na$_3$Sb and reported in the SM, Fig.~S7a, is analogous to its counterpart in K$_3$Sb. It stems from a transition at the $\Gamma$ point between the VBM, characterized by Sb $p$-contributions, and the CBm, where Na $s$-electrons reside~\cite{ette-degr00prb} (Fig.~S2b). 

Interestingly, P2 also originates from transitions near the $\Gamma$ point, involving the highest occupied and lowest unoccupied bands with analogous orbital characters as the VBM and CBm, respectively, see Fig.~S2a,b. The energy difference of approximately 0.5~eV between P1 in P2 in the spectra of both K$_3$Sb and Na$_3$Sb can be attributed to the steep parabolic band dispersion of the first conduction band around its minimum. In contrast, P3 exhibits a different character, visualized in Fig.~\ref{fig:exc_31}b for K$_3$Sb. It arises from numerous single-particle transitions distributed across a significant portion of the Brillouin zone including contributions from the high-symmetry point X, dominated by K $s$-states hybridized with Sb $p$-states in the CBm~\cite{kala+10jpcs}. This picture aligns with the larger oscillator strength of P3 compared to P1 and P2, suggesting a more localized nature of this exciton in real space. 

We now turn to the optical spectra of the hexagonal crystals with a 3:1 stoichiometric ratio, K$_6$Sb$_2$ and  Na$_6$Sb$_2$. Due to symmetry, the macroscopic dielectric tensor of these two materials exhibits two inequivalent optical components for in-plane ($\parallel$) and out-of-plane ($\perp$) light polarization (blue and red curves in Fig.~\ref{fig:os_31}c-d).
K$_6$Sb$_2$ and  Na$_6$Sb$_2$ have absorption onset in the near-infrared region, 0.68~eV for K$_6$Sb$_2$ and 0.92~eV for Na$_6$Sb$_2$, consistent with the trends of the respective QP gaps. Interestingly, in K$_6$Sb$_2$, the first absorption peak appears in the out-of-plane component and is thus labeled P1$_{\perp}$, see Fig.~\ref{fig:os_31}c, while in Na$_6$Sb$_2$, it is found in the in-plane polarization and marked as P1$_{\parallel}$, see Fig.~\ref{fig:os_31}d. The energy separation between the first maxima in the in-plane and out-of-plane components differs significantly in the materials: approximately 70~meV in K$_6$Sb$_2$ and 260~meV in Na$_6$Sb$_2$. A similar trend is noticed for the higher-energy peaks of varying intensity in the two spectra. For instance, P3$_{\perp}$ in Na$_6$Sb$_2$ is significantly more intense than P3$_{\parallel}$, which appears as a weak shoulder in Fig.~\ref{fig:os_31}d.

We can assess the binding energies of the excitations associated with these peaks using Eq.~\eqref{eq:Eb}. In the spectrum of K$_6$Sb$_2$, both P1$_{\perp}$ and P1$_{\parallel}$ are redshifted by 110~meV from their IQPA counterparts, while for Na$_6$Sb$_2$, the binding energy decreases to 90~meV (see Table~\ref{table:gaps}). The binding energies associated with P2$_{\parallel}$ and P2$_{\perp}$ reach 140~meV for both crystals while those of P3$_{\parallel}$ and P3$_{\perp}$ amount to 190~meV. Notably, in Na$_6$Sb$_2$, electron-hole correlations significantly impact the oscillator strength distribution, see Fig.~\ref{fig:os_31}d. This leads to strong optical anisotropy in the visible range, with enhanced absorption along the out-of-plane direction despite the lowest-energy excitation being in-plane polarized.

The composition of the excitations in the spectra of the hexagonal crystals is strikingly similar to their cubic counterparts, see Fig.~\ref{fig:exc_31}. The lowest-energy transitions occur between the VBM and the CBm, as illustrated in Fig.~\ref{fig:exc_31}c for P1$_{\perp}$ of K$_6$Sb$_2$. As the band edges retain the same orbital character as in cubic K$_3$Sb, namely Sb-$p$ at the VBM and K-$s$ at the CBm~\cite{ette-degr99jpcm} (see also Fig.~S2c), the nature of the exciton does remains consistent regardless of the polarization of the incoming light. Higher-energy excitations, such as P2$_{\parallel}$ and P2$_{\perp}$ in both K$_6$Sb$_2$ and Na$_6$Sb$_2$, involve electronic transitions slightly shifted from the $\Gamma$ point (see Fig.~S6b and S7b). As discussed above for the cubic crystals, such a strong localization in k-space mirrors a real-space delocalization. In contrast, P3$_{\parallel}$ and P3$_{\perp}$ originate from different states (see Fig.~\ref{fig:exc_31}d for K$_6$Sb$_2$ and Fig.~S7b for Na$_6$Sb$_2$). While the majority of transition weights are associated with electronic states near $\Gamma$, their distribution in k-space is broad, extending to the high-symmetry point A. These characteristics suggest a pronounced real-space localization of this excitation.

%%%%%
%
\begin{figure*}
	\centering
	\includegraphics[width=1.0\textwidth]{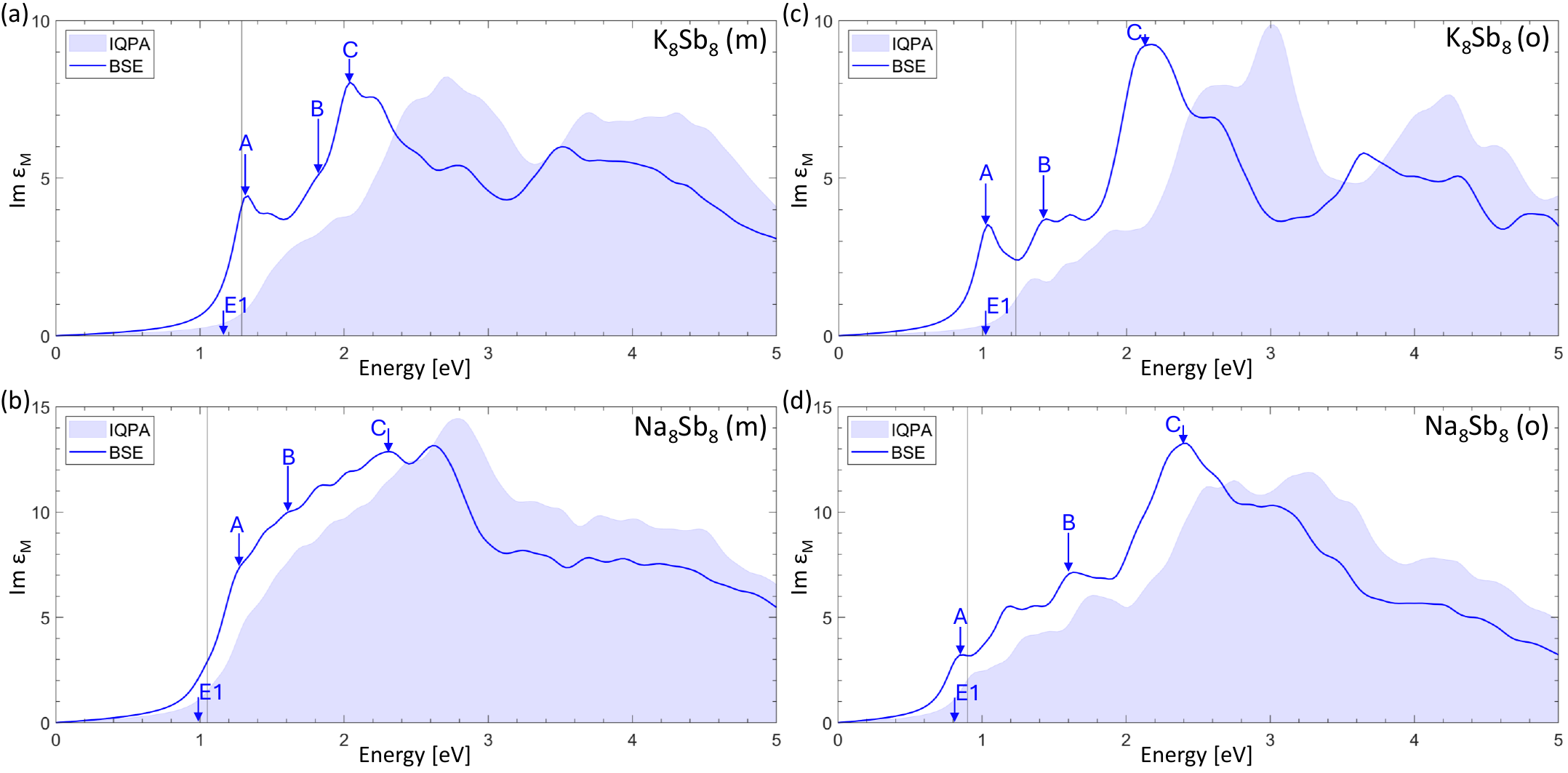}
\caption{Optical spectra of the monoclinic crystals (a) K$_8$Sb$_8$ and (b) Na$_8$Sb$_8$, as well as of the orthorhombic polymorphs (c) K$_8$Sb$_8$ and (d) Na$_8$Sb$_8$. The solid curves and the shaded areas represent the spatial average of the diagonal components of the imaginary part of the macroscopic dielectric tensor calculated from the solution of the BSE and in the IQPA, respectively. A Lorentzian broadening of 100~meV is applied to all spectra. In all plots, the QP gap is marked by a vertical bar.
}
	\label{fig:os_11}
\end{figure*}

We continue our analysis with the alkali antimonide crystals with a 1:1 stoichiometry, specifically the monoclinic and orthorhombic phases of K$_8$Sb$_8$ and Na$_8$Sb$_8$, see Fig.~\ref{fig:os_11}. Due to the lower symmetries of these lattices, the dielectric function exhibits three distinct diagonal components. Additionally, the monoclinic polymorphs possess a non-zero off-diagonal component visualized in the SM, Fig.~S5. To simplify the discussion, we plot in Fig.~\ref{fig:os_11} a spatial average of the diagonal components of the imaginary part of the macroscopic dielectric tensor computed for each material. The separate components are reported in the SM, Fig.~S4. These crystals are characterized by a broad absorption spectrum with less defined peaks compared to those with a 3:1 stoichiometric ratio, despite having a similar near-infrared absorption onset around 1~eV. This characteristic can be attributed to a higher density of states around the gap, see Fig.~\ref{fig:bandstructures}e-h and Ref.~\cite{schi+24ats}.

The spectrum of monoclinic K$_8$Sb$_8$ (Fig.~\ref{fig:os_11}a) is dominated by two prominent peaks, A and C, centered at 1.33~eV and 2.0~eV, respectively. A less intense peak, B, located at 1.82~eV, lies within a dense manifold of weaker excitations that cannot be individually resolved with the chosen Lorentzian broadening of 100~meV. We use a different notation for the peaks in these spectra compared to those displayed in Fig.~\ref{fig:exc_31} because the discussed features do not strictly correspond to the first, second, and third maxima. The lowest-energy excitation, E$_1$, is very weak and lies 120~meV below the QP gap (Table~\ref{table:gaps}). This behavior is expected due to the similar atomic and orbital character (Sb $p$) of the VBM and CBm~\cite{schi+24ats}. In this case, it is not straightforward to apply Eq.~\eqref{eq:Eb} to determine binding energies. Instead, we rely on the conventional definition of semiconductor physics~\cite{fox10book}, using the fundamental gap (in our case, the QP gap) as a reference. On this basis, none of the peaks marked in Fig.~\ref{fig:os_11}a corresponds to a bound exciton. However, a visual comparison of the BSE and IQPA spectra indicates an influence of electron-hole correlations on the spectral shape. Peak A, located a few hundred meV above the IQPA absorption edge, appears sharper and redshifted due to an oscillator strength redistribution. Similarly, peak C, which originates from a feature around 2.4~eV in the IQPA spectrum, is significantly redshifted and narrowed by electron-hole correlations in the BSE result. 

The spectrum of monoclinic Na$_8$Sb$_8$ is rather different, see Fig.~\ref{fig:os_11}b. The inclusion of excitonic effects in the BSE calculations primarily results in a rigid redshift by a few tens of meV of the broad spectral shape featured in the IQPA. The maximum near 3~eV experiences a pronounced redshift of 100~meV despite having higher oscillator strength in the IQPA spectrum. Similar to monoclinic K$_8$Sb$_8$, the lowest-energy excitation, $E_1$, is extremely weak and appears at 0.98~eV, namely 70~meV below the QP gap, see Table~\ref{table:gaps}. Peaks A, B, and C, are located at 1.28~eV, 1.63~eV, and 2.31~eV, respectively, and, as such they all lie above the QP gap.

The spectrum of orthorhombic K$_8$Sb$_8$ (Fig.~\ref{fig:os_11}c) exhibits similar characteristics to the monoclinic polymorph of the same material. The absorption onset is marked by peak A at 1.03~eV, coinciding with the lowest-energy excitation $E_1$. This relatively sharp maximum is followed by a weaker feature, B, at 1.44~eV. The broader peak C, centered at 2.15~eV, dominates the visible band. A comparison of the BSE and the IQPA spectra suggests that all three peaks have an excitonic nature. Although only peak A is a bound exciton with a binding energy of 200~meV (Table~\ref{table:gaps}), according to the conventional definition~\cite{fox10book} adopted for these crystals, both peak A and peak B are evidently redshifted from their IQPA counterparts and gain oscillator strength. Conversely, peak C and its neighboring shoulder undergo a mutual redistribution of spectral weight, similarly found in the electron energy loss near edge structure of complex materials like gallium oxide~\cite{cocc+16prb} and its alloys~\cite{balo+24acsaem}. 

Finally, the spectrum of orthorhombic Na$_8$Sb$_8$ exhibits several distinct peaks within a broad absorption band, see Fig.~\ref{fig:os_11}d. Peak A, centered at 0.84~eV, is slightly above the lowest-energy excitation $E_1$ at 0.81~eV, which is optically inactive. Hence, both peak A and $E_1$ are below the QP gap, and have a binding energy of 60~meV and 90~meV, respectively (Table~\ref{table:gaps}). Peak B, centered at 1.63~eV, has a corresponding local maximum in the IQPA spectrum, about 100~meV above it and with similar relative intensity. On the other hand, peak C at 2.43~eV gains more oscillator strength compared to the lower-energy peaks A and B when electron-hole correlations are taken into account.

\begin{figure*}
	\centering
	\includegraphics[width=1.0\textwidth]{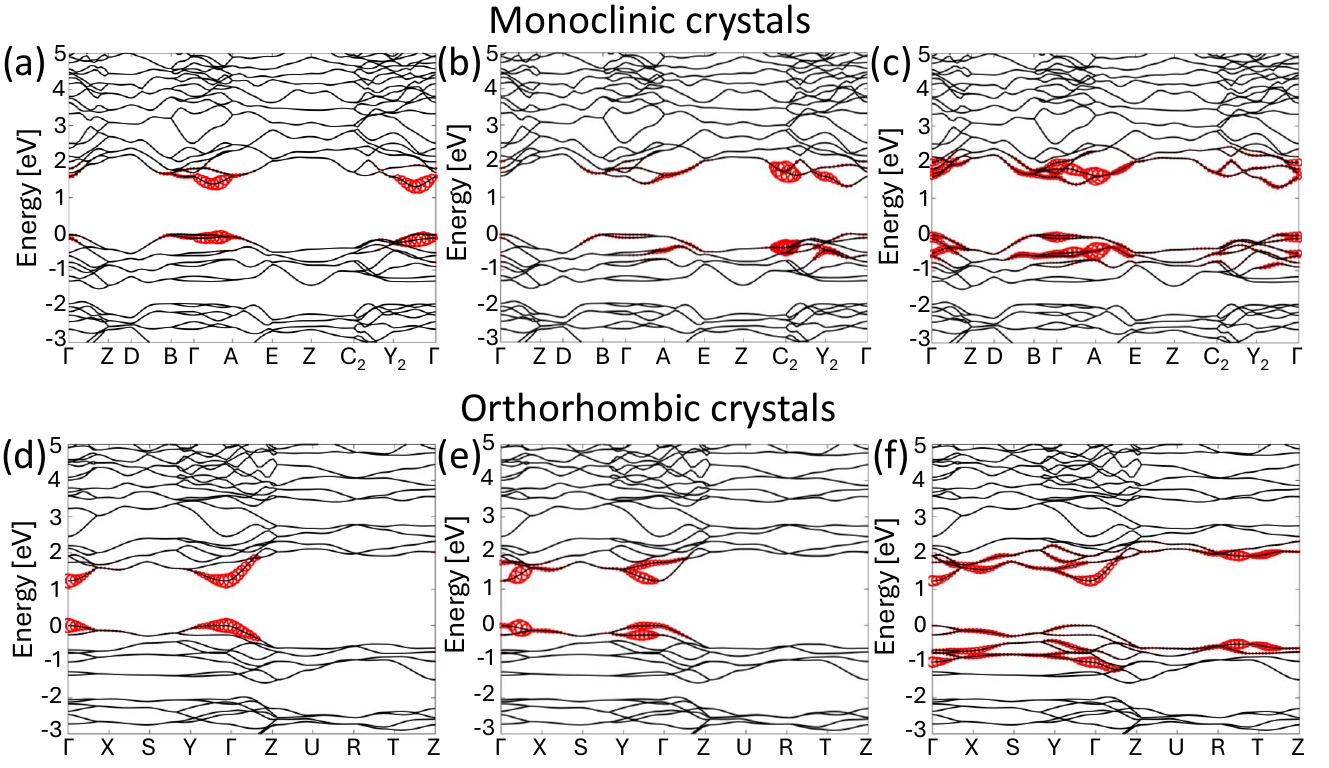}
\caption{Band-structure distribution of selected exciton weights, indicated by red dots of varying size, of the alkali antimonide crystals with 1:1 stoichiometric ratio. The results obtained for (a) peak A, (b) peak B, and (c) peak C of monoclinic K$_8$Sb$_8$, as well as (d) peak A, (e) peak B, and (f) peak C of orthorhombic K$_8$Sb$_8$, are visualized here. 
}
	\label{fig:exc_11}
\end{figure*}

The complex nature of the optical spectra of these materials is further reflected in the analysis of their excited states. In Fig.~\ref{fig:exc_11}, we display exemplary distributions of the weights associated with excitons A, B, and C of monoclinic (upper panels) and orthorhombic (lower panels) K$_8$Sb$_8$, with similar trends obtained for Na$_8$Sb$_8$, see SM, Fig.~S8 and S9. The excitation associated with peak A in the monoclinic crystal (Fig.~\ref{fig:os_11}a) originates from transitions near the VBM and CBm since it does not correspond to the lowest-energy excitation. The highest valence band and the lowest conduction bands have a predominant Sb $p$- and a hybridized Sb $sp$-character, respectively (Fig.~S3a,b). The weight distribution in k-space of this transition is relatively broad, involving the highest-occupied and the lowest-unoccupied bands along the B~$\rightarrow \Gamma \rightarrow$~A and Y$_2 \rightarrow \Gamma$ paths. 
The excitation associated with peak B is more scattered, involving states beyond the highest valence and lowest conduction bands, see Fig.~\ref{fig:exc_11}b. The largest weight concentration is around the high-symmetry point C$_2$ but non-negligible contributions come also from Y$_2$, where transitions from the fourth and fifth occupied bands target the lowest unoccupied one. It should be noticed that all these bands belong to the same manifold and span an energy range of less than 1~eV. Finally, the excited states forming peak C exhibit an even more complex composition, involving transitions from the four highest valence bands to the four lowest conduction bands across almost the entire Brillouin zone. 

The excitations marked in the spectra of the orthorhombic crystals (lower panels of Fig.~\ref{fig:exc_11}) feature similar trends but distinct characteristics. The lowest-energy excitation $E_1$, giving rise to peak A in orthorhombic K$_8$Sb$_8$ (Fig.~\ref{fig:os_11}c), originates from the VBM $\rightarrow$ CBm transition at $\Gamma$, see Fig.~\ref{fig:exc_11}d. The character of these states is analogous to the one discussed for the monoclinic crystals: a predominant Sb $p$-character in the top of the valence band and a hybridized Sb $sp$-nature of the lowest conduction band (Fig.~S3c,d). A similar scenario occurs in Na$_8$Sb$_8$, where the first excitation and the one forming peak A are energetically close, see SM, Fig.~S9b). The excitation associated with peak B arises from transitions involving the second occupied and unoccupied bands. They are primarily localized around the $\Gamma$ point and extend toward the high-symmetry points X and Y, where the uppermost valence bands and the lowest conduction bands are energetically closer (Fig.~\ref{fig:exc_11}e). Finally, peak C involves contributions from multiple valence and conduction bands up to the sixth occupied and fourth unoccupied states, respectively (Fig.~\ref{fig:exc_11}f). The k-space distribution of their weights is rather delocalized and involves states in the region between the high-symmetry points R and T, where the valence and conduction bands are separated by more than 3~eV. 
The isostoichiometric nature of these crystals, regardless of their lattice, influences the character of the electronic bands at the frontier. In contrast with the 3:1 compounds, here, the relatively larger ratio of Sb enhances its contributions, especially in the conduction region where unoccupied $p$- and $d$-orbitals are found closer to the gap (see Fig.~S3). The VBM retains instead its predominant Sb $p$-character as in Na$_3$Sb and Na$_6$Sb$_2$~\cite{schi+24ats}.

\section{Summary and Conclusions}
In summary, using DFT and MBPT, we investigated the electronic and optical excitations of eight binary K-Sb and Na-Sb crystals with 3:1 and 1:1 alkali-to-antimony stoichiometric ratios and diverse lattice symmetries. All but monoclinic Na$_8$Sb$_8$ exhibit direct band gaps ranging from approximately 0.8~eV to 1.3~eV. Optical spectra feature near-infrared absorption onsets and intense visible absorption bands. Excitonic effects are more pronounced in the K-based compounds, with binding energies exceeding 100~meV and sharp absorption peaks. In contrast, Na-based crystals exhibit binding energies below 100 meV and broader spectra, particularly for the monoclinic and orthorhombic polymorphs with the 1:1 stoichiometric ratio. The analysis of the electronic contributions to the excited states offers additional insight into the role of electron-hole correlations. While cubic and hexagonal crystals with 3:1 composition are dominated by low-energy Wannier excitons, in the isostoichiometric materials, the distribution of the excitons is more sparse in k-space, suggesting more pronounced localization in real space. 

In conclusion, our first-principles study provides valuable insights into the fundamental properties of binary alkali antimonide crystals. These materials can emerge during the growth of ternary compounds like Na$_2$KSb, which are gaining increasing attention as promising electron sources for particle accelerators~\cite{rozh+24prap}. Using state-of-the-art many-body perturbation theory, our analysis complements high-throughput DFT screening efforts aimed at identifying and characterizing materials with enhanced photoemission properties~\cite{anto+21am,schi+24ats}. The near-infrared absorption onset predicted for all considered compounds is particularly promising for this application. We anticipate that these findings will stimulate further experimental investigations of these promising alkali antimonide phases.

\section*{Acknowledgements}
This work was funded by the German Research Foundation (DFG), Project No. 490940284. C.C. acknowledges additional funding from the German Federal Ministry of Education and Research (Professorinnenprogramm III), and the State of Lower Saxony (Professorinnen f\"ur Niedersachsen). Computational resources were provided by the HPC cluster ROSA at the University of Oldenburg, funded by the DFG (project number INST 184/225-1 FUGG) and the Ministry of Science and Culture of the Lower Saxony State.

%%%%%%%%%%%%%%%%%%%%%%%%%%%%%%%%%%%%%%%%%%%%%%%%%%%%
\section*{Data availability}
The data that support the findings of this article are openly available: https://doi.org/10.5281/zenodo.14510458
%%%%%%%%%%%%%%%%%%%%%%%%%%%%%%%%%%%%%%%%%%%%%%%%%%%%%%%%

%\bibliography{bib}


\end{document}